\documentclass[aps,prc,article]{revtex4}%
\usepackage{amsfonts}
\usepackage{amsmath}
\usepackage{amssymb}
\usepackage{graphicx}%
\begin{document}
\title{Long-range multiparticle interactions induced by neutrino exchange \\in neutron-star matter}
\author{M. I. Krivoruchenko}
\affiliation{National Research Centre ``Kurchatov Institute'' \\ Ploschad' Akademika Kurchatova 1$\mathrm{,}$ 123182 Moscow, Russia}

\begin{abstract}
Forces with a large radius of interaction can have a significant impact on the equation of state of matter.
Low-mass neutrinos generate a long-range potential due to the exchange of neutrino pairs. We discuss a possible relationship between the neutrino masses, which determine the interaction radius of the neutrino-pair exchange potential, and the equation of state of neutron matter.
Contrary to previous statements, the thermodynamic potential, when decomposed into the number of neutrino interactions, vanishes in any decomposition order, except for the interaction of two neutrons. In the one-loop approximation, long-range multiparticle neutrino interactions are stable in the infrared region for all neutrino masses and do not affect the equation of state of neutron matter or the stability of neutron stars.
\end{abstract}
\maketitle

Among the fermions of the Standard Model, neutrinos are the lightest particles.
Their masses are at least six orders of magnitude smaller than the mass of any other charged fermion.
The exchange of low-mass particles creates a long-range potential.
The exchange of massless photons, e.g., leads to the Coulomb potential.
Since neutrinos are fermions, long-range two-body forces can involve them through the formation of neutrino pairs
\cite{Feinberg:1968, Hsu:1994, Grifols:1996, Fischbach:1996, Segarra:2020}.
The neutrino-pair exchange potential is similar to the van der Waals potential arising from the two-photon exchange (see, e.g., \cite{Itzykson:1980}).
Weakly interacting light bosons provide an excellent illustration
of the significant influence of weak forces with large interaction radii on the equation of state (EoS) of neutron matter and the structure of neutron stars \cite{Krivoruchenko:2009,Wen:2009}.

Fischbach \cite{Fischbach:1996} considered the effect of long-range multiparticle interactions of neutrinos
on the EoS of neutron matter and concluded that the contribution of neutrino interactions to EoS diverges
in the infrared region.
To guarantee the finiteness of EoS and, ultimately, the stability of neutron stars,
he postulated a lower limit for the neutrino masses of $m \gtrsim 0.4$ eV.
Cosmological models place an upper limit
on the sum of neutrino masses of 0.13 eV \cite{Abbott:2022}. According to
KATRIN experiment on tritium $\beta$ decay, the upper limit on the effective
electron neutrino mass is 0.8 eV \cite{KATRIN:2022}.
Fishbach's estimate is close to these limits and partly intersects with them,
which requires a thorough analysis of multiparticle neutrino interactions in nuclear matter.
In relation to the KATRIN experiment,
the mass constraint \cite{Fischbach:1996} is discussed in a recent paper \cite{Fischbach:2022}.

Abada, Gavela, and Pinea \cite{Abada:1996} 
address the effect of multiparticle neutrino interactions 
using the standard methods of quantum statistics (see, e.g., \cite{LL:1978}).
The authors confirm an infrared instability of EoS in each order of
decomposition by the number of interactions.
Abada \textit{et. al.} conclude, nevertheless, that the total contribution
of multiparticle neutrino interactions to the EOS of neutron matter is zero.

In this paper, we show that, contrary to the previous statements \cite{Fischbach:1996, Abada:1996},
the multiparticle interactions of neutrinos are stable in the infrared region and, moreover,
their contributions vanish at each term of the EoS expansion into a power series with respect to the number density.
The structure of neutron stars is thereby not sensitive to the mass of neutrinos.

The effective Hamiltonian for low-energy interaction of neutrinos and neutrons
is generated by the exchange of the Z boson.
We consider the case of Dirac neutrinos.
After averaging the neutral weak current
of quarks over the neutron wave function, the effective Hamiltonian takes
the form
\begin{equation}
H_{\text{eff}}=-\frac{G_{F}}{2\sqrt{2}}\int d^{3}x\left[  \bar{\nu}%
(x)\gamma_{\mu}(1-\gamma_{5})\nu(x)\right]  \left[  \bar{n}(x)\gamma^{\mu
}(1-g_{A}\gamma_{5})n(x)\right]  , \label{Hint}%
\end{equation}
where $g_{A}$ is the axial coupling constant of nucleons. The axial
component of the weak current, which indicates the direction of the
average spin of neutrons, vanishes in unpolarized matter, so the Z-boson mean field, $U$, is a pure vector.
The massive neutrino interacts with the potential $U$ by its left component only.
In mean-field approximation, $\langle\bar{n}(x)\gamma^{\mu}(1-g_{A}\gamma_{5})n(x)\rangle=g^{\mu0}\rho$.
The typical number density of neutrons is $\rho \sim 0.4$ fm$^{-3}$.
The typical Fermi momentum of neutrons is several hundred MeV, whereas the Z-boson
mean-field potential equals $U=-G_{F}\rho/\sqrt{2} \sim- 20$ eV.
We work in the approximations of homogeneous neutron matter and flat Minkowski space.
These approximations are well-founded for neutrinos with masses greater
than the inverse radius of neutron stars,
i.e. $m \gtrsim 1/R_{\mathrm{s}} \sim 2\times10^{-11}$ eV, where $R_{\mathrm{s}} \sim 10$ km.

It is useful to define projection operators $L=(1-\gamma
_{5})/2$, $R=(1+\gamma_{5})/2$, and $P_{\pm}=(1\pm
\mbox{\boldmath$\alpha$}\mathbf{{n})}/2$, where
$\mbox{\boldmath$\alpha$}=\gamma_{0}\mbox{\boldmath$\gamma$}$ and
$\mathbf{n}=\mathbf{q}/|\mathbf{q}|$ is the unit vector oriented in the
direction of neutrino momentum.

The effective Lagrangian of a neutrino with mass $m$ has the form:
\begin{equation}
\mathcal{L}_{\mathrm{eff}}(x)=\bar{\nu}(x)(i\hat{\nabla}-U\gamma_{0}%
L-m)\nu(x).
\end{equation}
The Green function is defined by the quadratic form of the effective
Lagrangian. In the momentum representation,
\begin{equation}
\hat{S}_{F}(q,U)=\frac{1}{\hat{q}-U\gamma_{0}L-m}. \label{SF}%
\end{equation}

The change in the thermodynamic potential, $\Omega$, due to an interaction
is given by the well-known expression (see, e.g., \cite{LL:1978})
\begin{equation}
\Omega-\Omega_{0}=\int_{0}^{1}\frac{d\lambda}{\lambda}\langle H_{\text{eff}%
}^{\lambda}\rangle. \label{DOmega}%
\end{equation}
In the case under consideration, $H_{\text{eff}}^{\lambda}=\lambda
H_{\text{eff}}$ is the effective Hamiltonian (\ref{Hint}) with the scaled
coupling constant. In terms of the Green function,
\begin{equation}
\Omega-\Omega_{0}=V\int_{0}^{1}\frac{d\lambda}{\lambda}\lim_{\tau
\rightarrow-0}\int\frac{d^{4}q}{(2\pi)^{4}}e^{-iq_{0}\tau}(-i)\mathrm{Tr}%
\left[  \lambda U\gamma_{0}L\hat{S}_{F}(q,\lambda U)\right]  ,
\label{DOmegaSF}%
\end{equation}
where $V$ is the normalization volume. This expression implies a smooth
thermodynamic limit $V \to \infty$, $\rho = \textrm{const}$. In a realistic approach, the number of particles in
a star is finite, although large.
The neutrino propagator should be expanded into a power series by the neutron number density, i.e.,
by the parameter $U$, and the series should be truncated at $s \sim N \equiv M_{\odot}/m_{n}=1.2\times
10^{57}$, where $M_{\odot}$ is the mass of the sun, and $m_{n}$ is the mass of
the neutron. Each term, proportional to $U^{s}$, describes the scattering of
neutrinos by $s$ neutrons. If the series converges, the limit $N
\to\infty$ is well defined and the decomposition is not required.

The papers \cite{Fischbach:1996,Abada:1996} declare that for massless neutrinos, the individual
terms of the series are proportional to $(UR_{\mathrm{s}})^{s}$.
If this were true, the infinite series would diverge because $UR_{\text{s}} \sim 10^{12}  \gg 1 $.
Abada \textit{et al.} further argue the transition to the limit of $N \to\infty$
by the possibility for neutrinos to scatter several times on the same neutron.
Multiple scattering involving the same neutron is possible only in higher orders of
the loop expansion, whereas Abada \textit{et al.} work in the one-loop approximation.
In this approximation, the transition to the limit of $N \to \infty$ is not allowed
if the series does not converge.

We expand the expression (\ref{DOmegaSF}) in a series by $U$. The
integration by $\lambda$ gives
\begin{equation}
\Omega-\Omega_{0}=V\lim_{\tau\rightarrow-0}\int\frac{d^{4}q}{(2\pi)^{4}%
}e^{-iq_{0}\tau}(-i)\mathrm{Tr}\left[  \sum_{s=1}^{N}\frac{1}{s}\left(
U\gamma_{0}L\frac{1}{\hat{q}-m}\right)  ^{s}\right]  . \label{DOmegaSFS}%
\end{equation}
The relations $LR=0$, $L\gamma_{\mu}=\gamma_{\mu}R$ lead to the identity
\begin{equation}
\mathrm{Tr}\left[  \left(  \gamma_{0}L\frac{1}{\hat{q}-m}\right)  ^{s}\right]
=\mathrm{Tr}\left[  \left(  \frac{\gamma_{0} \hat{q}}{q^{2}-m^{2}}\right)
^{s}R\right]  . \label{transform 1}%
\end{equation}

In terms of the projection operators defined above,
$\gamma_{0}\hat{q} =(q_{0}-|\mathbf{q}|)P_{+} +(q_{0}+|\mathbf{q}|)P_{-}$.
Using the binomial formula for $(\gamma_{0}\hat{q})^s$ and the relations $P_{+}P_{-}=0$, $P_{\pm}^{s}=P_{\pm}$,
the right side of Eq.~(\ref{transform 1}) can be simplified to give
\begin{equation}
\mathrm{Tr}\left[  \frac{(q_{0}-|\mathbf{q})^{s}P_{+}+(q_{0}+|\mathbf{q}%
|)^{s}P_{-}}{(q^{2}-m^{2})^{s}}R\right]  . \label{transform 2}%
\end{equation}
Closing the contour of integration by $q_{0}$ in the upper half of the complex plane, we find
that the integral is determined by the residues at $q_{0}%
=-\sqrt{\mathbf{q}^{2}+m^{2}} + i0$. Using the symbolic computing software
package Maple \footnote{{\footnotesize {https://www.maplesoft.com/}}}, it is
possible to find the residues and the corresponding integrals over the momentum
space. It turns out that all the terms $5\leq s \leq100$ of the series vanish
identically.
After regularization of the neutrino loop, the term $s = 2$
becomes finite, whereas the terms $s = 3,4$ vanish. For $s = 1$, the integral
over an infinitely distant contour in the upper half of the
$q_{0}$-plane cancels the contribution of the residue at $q_{0}=-\sqrt{\mathbf{q}%
^{2}+m^{2}} + i0$.

After performing the Wick rotation $q_{0} \rightarrow i\omega$ and assuming
that the limit $\tau\to-0$ can be interchanged with the momentum integral,
a more general proof can be offered.
The integration by $\omega$ goes within the
limits $-\infty<\omega<\infty$. The spherical coordinate system in the Euclidean
space $(\omega,\mathbf{q})$ is defined by $\omega=\eta\cos\alpha$, $q_{x}%
=\eta\sin\alpha\cos\beta$, $q_{y}=\eta\sin\alpha\sin\beta\cos\gamma$, and
$q_{z}=\eta\sin\alpha\sin\beta\sin\gamma$. The angles are restricted by
$0\leq\alpha,\beta\leq \pi$, $0\leq\gamma\leq 2\pi$. The absolute value of
momentum is $|\mathbf{q}|=\eta\sin\alpha$. The volume element is $d^{4}%
q=i\eta^{3}d\eta\sin^{2}\alpha d\alpha\sin\beta d\beta d\gamma$. The radial
variable $\eta$ takes values in the interval $(0,+\infty)$. After integration over the angles,
the expression (\ref{transform 2}) becomes
\begin{equation}
\frac{(-i\eta)^{s}}{(\eta^{2}+m^{2})^{s}}\int2\cos(\alpha s)\times\sin
^{2}\alpha d \alpha \sin\beta d\beta d\gamma=-\frac{16\pi(-i\eta)^{s}}{(\eta
^{2}+m^{2})^{s}}\lim_{\xi\rightarrow s}\frac{\sin(\pi\xi)}{\xi(\xi^{2}-4)}.
\label{transform 3}%
\end{equation}
All the terms in Eq.~(\ref{DOmegaSFS}) vanish for $5 \leq s < +\infty$, because
$\sin(\pi s)=0$ for integer $s$ and the integral in the $\eta$ variable
converges. There is $\infty\times0$ uncertainty for
the values $s=1, 3, 4$. The divergence in the radial integral is
eliminated by the regularization, in which case the terms $s=1, 3, 4$ similarly vanish.

The term $s=2$ is stable in the infrared region, as evidenced by the neutrino-pair exchange potential for zero neutrino mass
\cite{Feinberg:1968}: 
\begin{equation}
U_{\mathrm{nn}}(r) = \frac{G_{F}^{2}}{16\pi^{3}r^{5}}. \label{potential}%
\end{equation}

Additional contributions to the potential (\ref{potential}) arise  from loops involving charged fermions of the Standard Model at distances closer than the electron's Compton wavelength and from loops involving heavy bosons of the Standard Model at distances closer than the Z boson's inverse mass.

\vspace{2mm}

We examined the multiparticle contributions of the neutrino interactions to the thermodynamic potential.
The infrared divergences discussed earlier in the literature
are actually absent in each individual
term of the decomposition (\ref{DOmegaSFS}) and, thus, in the sum.
As a consequence, the limit $N \to\infty$ is
unnecessary, and keeping the number of particles large but finite
is sufficient to provide the required proofs.
All of the components in Eq.~(\ref{DOmegaSFS}) vanish for massless and massive neutrinos,
with the exception of the two-body interaction, which is negligible.
Thus, under the one-loop approximation, long-range multiparticle
interactions of neutrinos have no impact on the structure and stability of neutron stars.

\vspace{2mm}

The author sincerely thanks
the organizers of the ITEP Astrophysical Seminar
for drawing his attention to ref.~\cite{Fischbach:2022}
and the participants of the seminar for interesting discussions.

\end{document}